\documentclass [floatfix,  superscriptaddress, twocolumn, showpacs, aps, pra, 10pt]{revtex4-1}

\usepackage{amsmath}
\usepackage{hyperref}
\usepackage{graphicx}
\usepackage{bbm}
\usepackage{amssymb}
\usepackage[english]{babel}
\usepackage{lmodern}
\usepackage[T1]{fontenc}
\usepackage[utf8]{inputenc}
\usepackage{wasysym}
\usepackage{amssymb}
\usepackage{gensymb} 
\usepackage{array} 

\graphicspath{{Figs/}}

\begin{document}

\title{Fast, cheap, and scalable magnetic tracker with an array of magnetoresistors}







\author{Valerio Biancalana} 
\affiliation{Dept. of Information Engineering and Mathematics - DIISM, University of Siena -- Via Roma 56, 53100 Siena, Italy}
\email{valerio.biancalana@unisi.it}
\author{Roberto Cecchi} 
\affiliation{Dept. of Physical Sciences, Earth and Environment - DSFTA, University of Siena -- Via Roma 56, 53100 Siena, Italy}
\author{Piero Chessa}
\noaffiliation
\author{Giuseppe Bevilacqua}
\affiliation{Dept. of Information Engineering and Mathematics - DIISM, University of Siena -- Via Roma 56, 53100 Siena, Italy}
\author{Yordanka Dancheva}
\affiliation{
Aerospazio Tecnologie srl, Strada di Ficaiole, 53040 Rapolano Terme (SI), Italy}
\author{Antonio Vigilante}
\affiliation{Department of Physics and Astronomy, University College London, Gower Street, London WC1E 6BT, United Kingdom}

\begin{abstract}
{We present the hardware of a cheap multi-sensor magnetometric setup, where a relatively large set of magnetic field components is measured in
several positions by calibrated magnetoresistive detectors. 
The setup is developed to map the (inhomogeneous) field generated by a known magnetic source, which is measured and then discerned from the background (homogeneous) geomagnetic field. The data output from this hardware can be successfully and reliably used to retrieve the position and  orientation of the magnetic source with respect to the sensor frame,  together with the orientation of the frame with respect to the environmental field. Possible applications of the setup are briefly discussed, and a synthetic description of the methods
of  data elaboration and analysis  is provided.}
\end{abstract}

\maketitle

\section*{\label{sec:introduction}Introduction}

\begin{figure}[ht]
   \centering
    \includegraphics [angle=00, width=  \columnwidth] {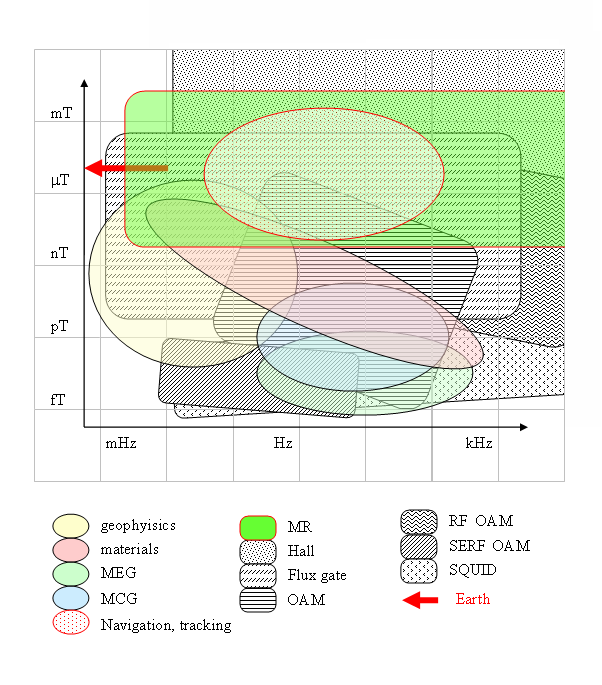}
  \caption{Indicative and approximate representation of typical field and frequency ranges for different kinds of magnetometry technologies, with their possible applications. Magnetoresistors 
  match requirements for navigation and tracking, while higher sensitivity is needed for biomagnetic detection (as in magneto-cardiography (MCG) and magneto-encephalography (MEG)), where SQUIDs and OAM are the eligible choices. OAM in their radio-frequency (RF) implementations offer excellent performance at high frequencies. Fluxgate sensors are produced in different grades of performance, and partially overlap with the MR applicability. Hall sensors are relegated to evaluating stronger fields, with quite limited precision.   }
   \label{fig:ranges}
\end{figure}

Magnetic field measurements can be performed with a 
 { variety of sensors characterized by very broad ranges in terms of sensitivity, robustness, dynamicity, linearity, reliability, speed, simplicity, and cost.}
The state-of-art sensors in terms of sensitivity are based on the superconductor quantum interference devices (SQUIDs),  which surpass the sensitivity level of $1\mathrm {fT}/\sqrt{\mathrm {Hz}} $.  The main drawback of SQUIDs is their need for cryogenics. Optical atomic magnetometry constitutes another technology that in some implementations -- particularly in the so-called Spin-Exchange-Relaxation-Free (SERF)-- may compete with SQUIDs in terms of sensitivity. It also enables  the construction of relatively simple and robust sensors with sensitivity at the $\mathrm {pT}/\sqrt{\mathrm {Hz}} $ level and below, including miniaturized devices, and implementations with a high-frequency response. 
Optical atomic magnetometers (OAMs) do not require cryogenics. On the other hand, they are commonly based on spectroscopy in high quality vapor cells illuminated with laser sources, both features that render them expensive and not easily integrable in solid state devices.
When these extreme performances are not required, fluxgate
technology offers an eligible alternative, on the basis of which different grades of sensors are produced with a rather wide range of sensitivities and costs. When sensitivities of the order of $\mathrm {nT}/\sqrt{\mathrm {Hz}} $ are sufficient, beside the low-cost fluxgate sensors,  solid-state technology now offers  extremely cheap and easy-to-use devices based on the  magnetoresistive effects. The magnetoresistive effect (discovered and first studied by Lord Kelvin \cite{ thomson_bak_56, thomson_roy_57})  in its giant \cite{baibich_prl_88},  tunnel \cite{yuasa_jpd_07}, and anisotropic \cite{mcguire_ieee_75} occurrences, can be profitably used to measure fields of several tens of micro Tesla, i.e. of the order of the geomagnetic field, with typical sensitivities in the  $\mathrm {nT}/\sqrt{\mathrm {Hz}} $ range and bandwidth extending up to kHz, or  --depending on the implementation-- to MHz and beyond \cite{tawfik_ieee_18}. Devices based on magneto-resistance
(MR) with sensitivity levels attaining  $\mathrm {pT}/\sqrt{\mathrm {Hz}} $ above 1~kHz have been reported \cite{liou_ieee_11}. The magnetoresistive sensors (similarly to SQUIDs and fluxgates, and in contrast to typical OAM magnetometers) are vectorial in nature, i.e. respond to single components of the magnetic field: an assembly of three orthogonally oriented sensors enables complete measurement of the magnetic field vector. Other popular solid state sensors are based on the Hall effect: their sensitivity and accuracy is worse than MR and they are more commonly used to roughly estimate (or rather just to detect) relatively strong fields.
A  synthetic, rough overview of the technologies available  and of the accessible bandwidth and sensitivity intervals for magnetometric measurements is provided in Fig.\ref{fig:ranges}.

The success of MR technology is also related  to its easy integration with silicon-based electronic devices. Integrated circuits (ICs) containing one or more magnetoresistive 
elements  {have recently become} very popular, and their cost has decreased by orders of magnitude over the last decade, thanks to  large-scale production: nowadays  two- or three-axis devices are widely used as sensors for electronic compasses, such as those contained in  smartphones and  drones. Other applications for these cheap sensors include virtual and augmented realities, navigation, non-destructive evaluation, and various industrial activities \cite{tawfik_ieee_18, jander_spie_05, jogschies_sens_15}.
Modern MR ICs contain not only magnetoresistive sensors but also  signal conditioning and DAQ electronics, together with special reset circuits that facilitate the MR use and improve reliability and reproducibility of their response, as well as circuitry for digital data transfer.

We  {focused} our attention on a family of ICs designed for Inter-Integrated-Circuit  (I$^2$C) protocol transmission of 3D magnetic data, which allows for an acquisition and data transfer rate as fast as 200 readings (3 data per read) per second. Due to the typical presence of a single sensor per user, in most cases these ICs have a static (not re-configurable) I$^2$C address. Some devices  {do, however, have a}  re-configurable address. In the case of the IC used in our prototypes, two address configuration pins (for a maximum of four chips per bus) are available. This limitation leads us to develop control-interface circuits with a parallelized architecture, when a sensor array is needed. At the cost of the obviously heavier circuitry, such an approach brings the second (but not secondary) advantage of accelerating the data acquisition and making it simultaneous  {throughout} the sensor set.

The aim of our work is to acquire, at a relatively high rate (hundreds  {of} samples per second), magnetometric data sets that can be elaborated to reconstruct the position of magnetic field sources, i.e. to track their spatial co-ordinates and the angles of their orientation. In the frequent case of a simple source like  
a magnetic dipole \cite{takaaki_ieee_06, hu_ieee_10,  bhashyam_spie_14}, the tracking procedure  provides three spatial coordinates and three components of the magnetic dipole. 
As soon as the measurement is performed in an external field (such as the geomagnetic field, which can be assumed homogeneous over the volume of the sensor frame), three more field components have to be worked out,  {making} a total of $6+3=9$ tracking data to be extracted per measurement. These 9 data contain redundant information in the case of repeated measurements, because both the modulus of the dipole moment and the modulus of the ambient field can be assumed to be constant. 
 {This means} that rotations of the dipole around its direction and rotations of the sensors around the ambient field do not cause field variations: these rotations correspond to \textit{blind co-ordinates}.  
In other terms, if it can be assumed  that the intensity of the magnetic source and/or of the environmental field are constant, the number of freedom degrees (and hence of the fitting parameters) would be reduced from 9 down to 8 or 7.

The information about the two blind co-ordinates could be retrieved  using a non-dipolar source (e.g. a set of two rigidly connected dipoles) and by complementing the environmental magnetic field with a measurement of the gravitational field. 
 {We will not address these possible improvements in this paper. } 
The use of non punctual sources, with the introduction of multipolar terms that break the dipole symmetry, has been successfully attempted and reported in the literature \cite{meng_ieee_09}.

Similarly, the system could be used to track two or more dipolar sources,  {located arbitrarily}  with respect to each other \cite{meng_ieee_10, meng_ieee_16}. In this case, the  enhancement in the source's degrees of freedom would require more computing resources and an increase of the minimum number of sensors.

The  {ability to track} objects with an adequately fast time response is a challenging and intriguing achievement, with important implications in several areas of research and applications. In particular, tracking methods based on magnetometric measurements offer a minimally invasive methodology and have been studied/proposed in a variety of applications, including medical diagnostics \cite{ dinatali_ieee_13}, the tracking of vehicles \cite{wahlstroem_ieee_14}, biology \cite{jun_pone_13}, and robotics \cite{than_ieee_12}. 

Further possible applications may arise in   {other diverse} areas, also depending on the precision and the speed that can be achieved, such as body part tracking (eye, tongue, hand, finger), human-computer interfaces, virtual and augmented reality, etc.

Several approaches have been proposed to  {deal with} the inverse problem of reconstructing the field source parameters from the field measurement (see e.g. ref. \cite{birsan_ieee_11} and references therein). Depending on the application,  {the requirements in terms of} accuracy, precision, robustness and speed of the tracking procedure may change, and different methodologies can be applied.

\section{Setup overview}
We have designed and built an interface circuit capable of operating arrays made of up to eight three-axis MR sensors, which can be differently disposed and grouped  {within} the space.
\begin{figure}[ht]
   \centering
    \includegraphics [angle=0, width=  \columnwidth] {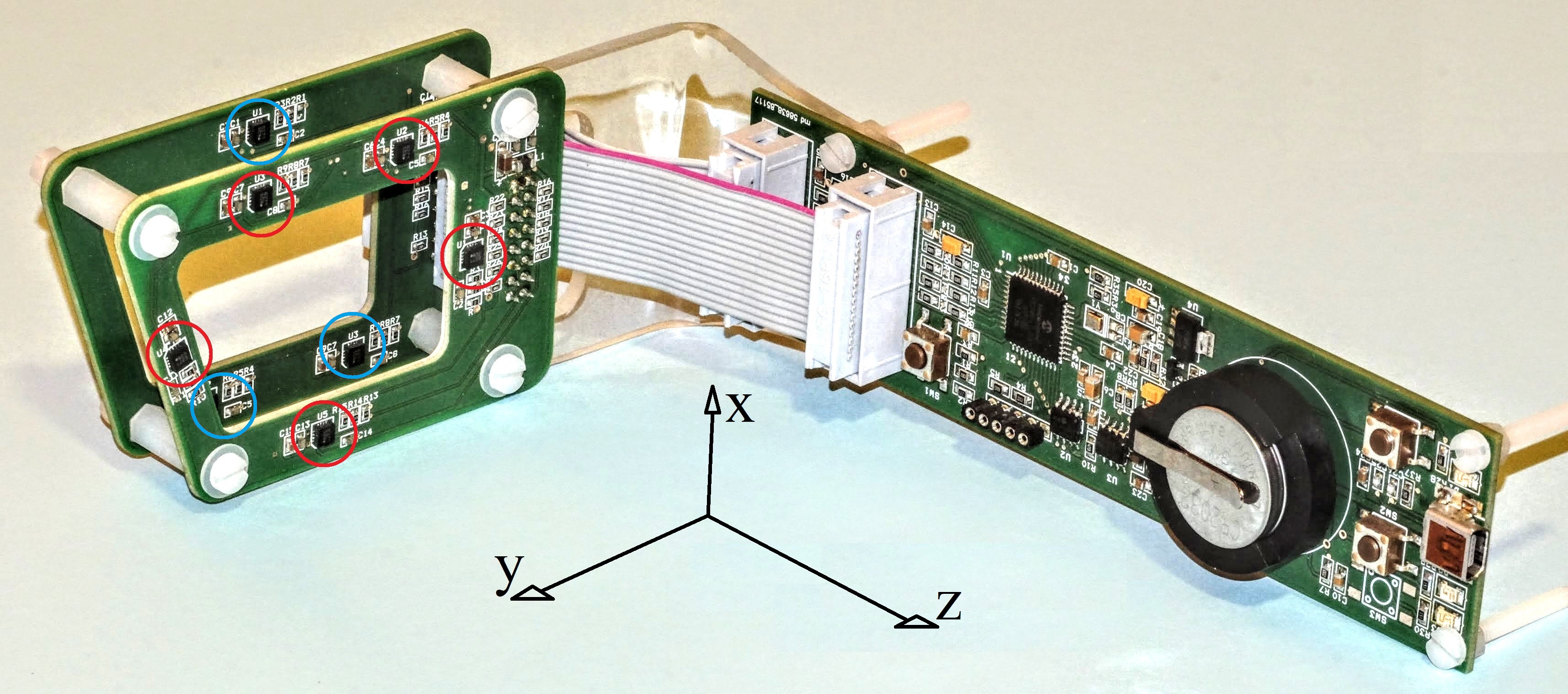}
  \caption{The prototype of the  electronics equipped with a sensor array designed to perform eye tracking. The eight three-axial sensors are mounted on the two parallel PCBs
  on the left: those on the front PCB are marked with red circles and those on the rear PCBs with blue circles. The acquisition, recording and data-transfer electronics is implemented in the  PCB 
  on the right, where the batteries and USB connector are visible. The flat connection make it possible to interface the PCBs
  with 3+5 or 4+4 sensor arrays.
  } \label{fig:setup}
\end{figure}
A picture of a sensor array and  {related} interface electronics is shown in Fig.\ref{fig:setup}. A printed circuit board (PCB) hosting a micro-controller,  Electrically Erasable Programmable Read-Only Memory (EEPROM), internal and external supply circuitry and a USB interface is connected to two  {sensor-PCBs}, which can host 4+4 or 5+3 IC sensors (as shown in that Fig.\ref{fig:setup}). Each sensor has a dedicated I$^2$C bus for communicating with the micro-controller. 
The limited number of address-bits available makes  some degree of parallelization necessary. We opted for a fully parallel architecture (one I$^2$C bus for each IC) that --at the expense of a slightly heavier circuitry-- brings several advantages, among which:  simultaneous measurements; simpler firmware; and no need to transmit data over the bus along the conversion (this avoiding a possible source of noise). Hence the microcontroller may send commands (and receive data) simultaneously to (and from) the eight sensors.
The data can be either immediately transmitted to the computer or stored 
on
an EEPROM  for  {subsequent} download. The second option is used  {when acquiring data for the} calibration procedure (see Sec.\ref{sec:calibration} below). Details of the  functionalities available and consequent possible operations are provided in the next sections. These are designed in view of producing fast (real time) tracking devices with a high throughput rate (100 trackings per second in the current implementation) and  precision (sub-millimetric spatial- and 2 degree  angular-resolution).

\section{Sensor specifications}
\label{sec:sensor}
The  Isentek  IST8308  IC \cite{isentek} is a three-axis magnetometer based on the anisotropic magnetoresistance effect, whose main characteristics are reported in Tab.\ref{tab:8308}. The chip implements reset, temperature compensation and analogue-to-digital conversion circuitry, and communicates trough an I$^2$C port. Both  on-demand and continuous data outputs can be queried. The maximum data acquisition rate is  200~Sa/s, and different kinds of internal filters can be activated to improve the  signal-to-noise ratio at the expense of the 
 {effective} bandwidth. Sensitivity and offset have non-negligible deviations from ideality  {(zero offset and maximum count at full-scale field),} so that a calibration is necessary (see Sec.\ref{sec:calibration}). These parameters may also change with time and (slightly) with temperature, so that an adequate measurement accuracy can be  {maintained} at the cost of repeated calibrations. It is worth noting that the temperature compensation circuitry  {helps reduce this problem,} since the effects of typical daily environmental temperature drifts can be neglected. In our implementation, a temperature sensor has been included, so that the user knows both the temperature at which the calibration data were collected and the current temperature:  when the deviation exceeds a threshold level, an alert is provided, making 
 {renewed} calibration advisable.
A data-consistency analysis is also available  {in order to highlight the need to re-calibrate} the unit, as discussed at the end of Sec.\ref{sec:calibration}.

\begin{table}
\centering
\begin{tabular}{ | m{4.8cm} | m{1.8cm}| m{1.2 cm} | } 
\hline
\bf parameter & \bf value & \bf unit \\ 
\hline
sensitivity drift with temperature & 0.023 & \% / $\degree$C \\ 
\hline
zero-field offset  & 300 & nT\\ \hline
offset drift with temperature & 27 & nT/ $\degree$C \\ 
\hline
dynamic range & 500 & $\mu$T \\ 
\hline
linear response & 200 & $\mu$T \\ 
\hline
nonlinearity/FS < & $10^{-3}$ & - \\ 
\hline
range (FS) & 200 or 500 & $\mu$T \\ 
\hline
hysteresis/FS < & $10^{-3}$ & - \\ 
\hline
supply voltage & 3.3 & V \\ 
\hline
size & $3\times 3\times 1$ & mm$^3$\\
\hline
adc resolution & 14 & bit\\
\hline
max output data rate & 200 & Sa/s\\
\hline
max I$^2$C clock & 400 & kHz\\
\hline
\end{tabular}
\label{tab:8308}
\caption{The IST8308 IC main features and characteristics.}
\end{table}

\section{Scalability}
\label{sec:scalability}
The system developed aims to  {simultaneously measure}  the environmental field and the field generated by a small, closely located magnetic source (dipolar magnet). The finite dynamic range of the sensors makes it necessary to deal with field contributions of comparable intensities. This condition should be fulfilled at least on a sensor subset providing a number of independent data sufficient to localize the magnet, i.e. not smaller than the number of freedom degrees of the system.

Good operating conditions can be identified as those in which the dipole  generates fields on the sensors ranging from 1/10 to 10 times the ambient one. Thus, keeping in mind the typical environmental field  value of some tens of $\mu$T, the $1/r^3$ dependence of the dipolar field, and the  magnetization values of permanent magnets (about 1T/$\mu_0$ for the  neodymium devices used), one finds that the best condition is fulfilled when the sensor-magnet distance is about 50 times larger than the linear size of the magnet. For example, a one cm$^3$ magnet produces a field comparable to the Earth's field at about a half-meter distance.  The chip size being sub-millimetric, this rule of thumb  applies when scaling down the system size as long as the sensor-magnet distance  remains much larger than the sensor size (e.g., magnets as small as 1~mm in size can be used at a distance of about 5~cm from the sensors).

Of course, according to the  tracking accuracy required, the sensor positions must be known with an adequate level of precision. In our case, the latter is determined by the PCB mount and is  sub-millimetric. However,  determination of the sensor positions can be improved on the basis of magnetometric data analysis \cite{meng_ieeeicrb_09},  and in some of our preliminary prototypes this kind of procedure was found to be crucial to guaranteeing the reliability of the tracking algorithms.

\section{Sensor calibration}
\label{sec:calibration}
MR sensors have a relatively accurate response in terms of linearity, but suffer from  {significant} offset and variable gain. Moreover, both gain and offset may vary with changes in temperature and  in time. In addition, unpredictable values are found after powering up the device. 

An important improvement has been introduced on the basis of a pulsed field-cycling technique. This reset field pulse technique  has also been  studied to improve the ultra-low frequency performance of MR devices \cite{he_rsi_11}. Modern magnetoresistance ICs contain apposite inputs to apply reset pulses (strong current pulses flow into a conductor built in the proximity of the magnetoresistive element, in order to re-magnetize its components in an appropriate and reproducible way). 

In more integrated devices the current pulses are produced internally  and the reset field cycle is automatically applied at the restart. The reset operation leads the sensor to work properly (reasonably low offsets and reasonably ideal gains along all the axes); nonetheless whenever good accuracy is required, some sort of calibration procedure \cite{merayo_mst_00} is advisable or necessary. In fact, the final offset value is substantially non-zero, and the gain may differ by several percent among the sensors contained in a single IC. An accurate evaluation of the offsets and gains makes it possible to convert the raw data into field measurements, so as to overcome these non-idealities. 

Our setup includes apposite parts devoted to facilitate the calibration procedure in the hardware, firmware, and software. Similarly to what is described in Ref.\cite{merayo_mst_00}, the calibration
procedure is accomplished by recording many data (simultaneously for all the 3D sensors of the array), while rotating the system freely and randomly in a (nominally) homogeneous field.  {In this measurement, each sensor detects} the magnetic field 
vector $\vec B$  as it moves on a spherical surface centered in the zero-field point of a Cartesian co-ordinate system: whenever the quantities  $\vec V$ do not span a spherical surface,
this can be due to non-zero offsets, to unequal gains, and to non-linear terms in the sensors' response. In the hypothesis of a linear response, the  quantities measured describe an off-center ellipsoid rather than a centered spherical surface: the displacement of the center is caused by the offsets, while the eccentricity is due to the gain anisotropy. 

It is worth noting that literature reports similar calibration procedures based on static measurements \cite{zikmund_ieee_15}. It is indeed possible to build a tri-axial field generator, finely calibrate it with the help of a scalar magnetometer and then use it to produce a rotating field of an assigned intensity. The latter can, in turn, be applied to calibrate vectorial sensors such as tri-axial magnetoresistive devices.

An optimization procedure is used to determine offsets and gains for all the sensors, and to save those values for subsequent data conversion. As described below (see Sec.\ref{sec:battery}), this kind of calibration measurement is more favorably performed with no cable connection. The optimization is usually done over data sets containing several hundreds of measurements (a maximum of N=2000 measurements is set by the EEPROM  size), with the minimized quantity being

\begin{equation}
\label{eq:f}
 f=\sum_{i,j,k, n} \left[ B_0^2- \left (T^{(k)}_{ij}\left(V_{j,n}^{(k)}-O^{(k)}_j \right) \right )^2 \right ]^2,  
\end{equation}
where  $V_{j,n}^{(k)}$ are the raw data; the indexes $i, j$ 
both run from 1 to 3 and denote the three Cartesian components; $k$ is the sensor index running from 0 to $K-1$, $O^{(k)}_j$ are the offset vectors to be determined; $T^{(k)}_{ij}$ are elements of triangular matrices (to be determined); $n$ denotes the measurement index (running from 1 to N); and $B_0$ is the ambient field modulus. The latter can be assigned arbitrarily, or from an independent measurement performed by scalar sensor (e.g. an atomic magnetometer) if quantitative field and dipole-moment estimations are required.

The elements $T^{(k)}_{ij}$ are ideally equal for $i=j$ (the inverse gains), while the off-diagonal elements are ideally zero for $i > j$. 
In contrast, the diagonal elements can be different from each other if the gain is not isotropic, and the off-diagonal elements could be non-zero, in the case of possible small misalignments (imperfect orthogonality) of the three axes. 

Let the optimal offsets and conversion matrices be given by $O^{(k-opt)}_j$ and $T^{(k-opt)}_{ij}$, respectively, and let's define 
\begin{equation}
    B^{(k)}_{i,n}=\sum_j T^{(k-opt)}_{ij} (V_{j,n}^{(k)} - O^{(k-opt)}_j)
\end{equation}
that is the $i^{th}$ component of the field in the position $\mathbf{r}_k$ of the $k^{th}$ sensor, as obtained from the sensor's output $V_j$ in the $n^{th}$ measurement.

Once the offsets are removed and the response has been made isotropic, further calibration is necessary to refer all the sensors  to one co-ordinate system. To this  {end}, the data recorded in the  calibration measurement mentioned are compared to each other. One sensor (let it be the $0^{th}$ one) is selected as a reference one,  and a rotation matrix is determined for each  sensor with $k=1... K-1$ by minimizing the vector differences between the field measured by that sensor and the reference one. In our implementation the rotation matrices are defined in terms of Euler angles, and the minimized quantities are:
\begin{equation}
   \label{eq:g}
    g=\sum_{i, j, n} \left( B_{i,n}^{(0)}- R_{i,j}^{(k)} B^{(k)}_{j,n} \right)^2,
\end{equation}
where $R_{i,j}^{(k)}=R_{i,j}(\theta^{(k)}, \phi^{(k)}, \psi^{(k)})$ are rotation matrices defined by the three angles $\theta^{(k)}, \phi^{(k)}, \psi^{(k)}$ to be determined for each $k = 1 ... K-1$.

In conclusion, each of the K sensors requires the determination of nine parameters (three offsets, three gains, three orthogonality-imperfection-compensation terms) for the conversion of the recorded values $\vec V^{(k)}$ into magnetic induction vectors 
$\vec B^{(k)}$  and each sensor (apart from the reference one) requires the determination of three rotation angles. The whole set of $m=9K+3(K-1)$ calibration parameters ($m=93$  in the  case considered, of $K=8$ sensors) is saved at the end of a calibration procedure and made available to perform the $\vec V \rightarrow \vec B$ conversions in the subsequent measurements.

Both the  minimizations of $f$ (eq.\ref{eq:f}) and $g$ (eq.\ref{eq:g}) are not critical in terms of convergence, and the calibration result is systematically reliable. Different algorithms can be used in order to execute the two tasks. In the current software implementation we are using the simplex (Nelder Mead) \cite{nelder_cj_65} 
routine, which is available among the Labview 
libraries.

Once the calibration parameters have been determined, a field estimation referenced to a unique Cartesian frame is available. This enables an additional procedure to check the validity persistence of the calibration parameters. This validation is performed by comparing the field components measured by the $K$ sensors in the homogeneous field with their median value. In particular, for each sensor, the software evaluates the quantity
\begin{equation}
 E_k=\frac{1}{N-1}\sum_{n=1}^{N} \sum_{i=1}^{3}  \left (B_{k,n,i}-\tilde B_{n,i} \right )^2,
\end{equation}
where $B_{k,n,i}$ denotes the $i^{th}$ component of the field as measured by the $k^{th}$ sensor at the $n^{th}$ measurement in a set of $N$, and $\tilde B_{n,i}$ denotes the median value of the $i^{th}$ component of the field measured by the $K$ sensors at the $n^{th}$ measurement. The presence of anomalously large $E_k$ values produces an alert, and the user can disregard the data from the corresponding sensor(s) in the subsequent tracking, or decide to  {perform} a new calibration.

\section{Power supply}
\label{sec:battery}
The circuit is normally supplied through the USB port, however it is possible to start it in a battery-supply mode, in order to acquire and store the calibration data with no cable constraints. To this end, there is a button to connect the battery, and a button to start the calibration measurement. When the calibration measurement starts, a circuit maintains the battery connection. During this self-supplied operation, a flashing LED denotes acquisition. At the end of the acquisition, the self-supplied mode is maintained for a 30 s, during which the operator can connect the cable. In this manner, the circuit remains supplied, and no reset pulses are applied. This feature is designed to guarantee that the data acquired  accurately describe the sensor response, since the latter could be modified in the case of an IC reboot, due to the  automatically applied reset pulses.

\section{Parallel I$^2$C buses}
\label{sec:i2c}
Apart from the  {above mentioned} problem arising from the fixed I$^2$C address (a feature that characterizes many types of MR ICs), it is advantageous to parallelize the communication with sets of ICs, both to accelerate the global data acquisition rate and to enable simultaneous (i.e. mutually time-consistent) measurements. 
We have studied and developed a simple but effective circuit, enabling both parallel data reading from the sensors and fast composite data transmission to a PC. The electronics developed  may communicate (for hardware configuration and  data transfer purposes) with eight ($K=8$) chips at once, thus providing  up to $K \times 3$ magnetometric data per reading. The data  transfer rate is limited either by the sensor throughput rate over the I$^2$C bus or by the composite data transmission rate over the USB port: in the present implementation a rate as  {high} 
as 100~Sa/s (2400~data/second) has been demonstrated, while  200~Sa/s (4800~data/second) is the limit set by the IC specifications.
Concerning the USB communication, this represents a potential bottle-neck. Its speed is  machine-dependent and may vary unpredictably e.g. with the number of processes running in the computer, and particularly with the presence of other connected interfaces.

\section{Firmware}

\begin{figure}[ht]
   \centering
    \includegraphics [angle=00, width=  \columnwidth] {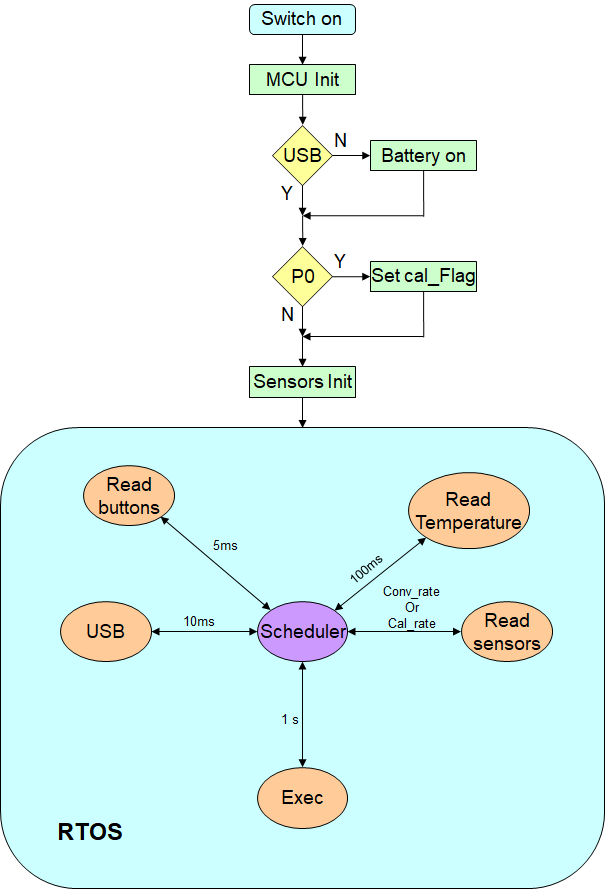}
  \caption{General flowchart of the firmware.
  } \label{fig:fwscheme}
\end{figure}

An overview of the firmware principle of operation is represented in Fig.\ref{fig:fwscheme}.
When the system starts up, the peripherals of the microcontroller are configured and the variables used (MCU Init) are initialized. In particular, among the peripherals configured, it is worth mentioning the ADC converter which measures the supply voltage and the timer for the real time operating system (RTOS) described below. In addition, a map is built of the sensors that are actually connected. 
The latter is then used when the data are transferred from the parallel I$^2$C buses to the on-board memory or to the USB interface, as described in Sec.\ref{sec:i2c}.

Subsequently, a test is performed to verify  {whether} the system has been started by connecting the USB cable or by pressing the power button: in the second case a switch (MOSFET transistor) is closed,  {in order to} maintain the battery power supply when the power button is released by the user.

At the same time, whether or not the system calibration button (P0) was also pressed during the power-up phase it is also evaluated; in this case a flag variable called cal\textunderscore Flag is set to True. This flag is used during the operation of the RTOS. 
The next operation (Sensor Init) is the  initialization of the sensors that have been detected and included in the sensor map.

Now the RTOS can start. A scheduler establishes the execution sequence and times of the various programs (tasks) within an iterated cycle. During this cycle  {the scheduler evaluates} --for each task-- whether it is time to run it, on the basis of the time elapsed after the last execution: individual time intervals are defined for each task.

\begin{figure}[ht]
   \centering
        \includegraphics [angle=00, width=0.45 \columnwidth] {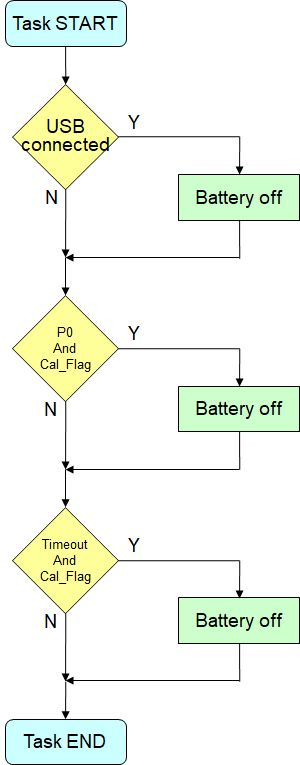}
        \hspace{1.5cm}
        \includegraphics [angle=00, width=0.35 \columnwidth] {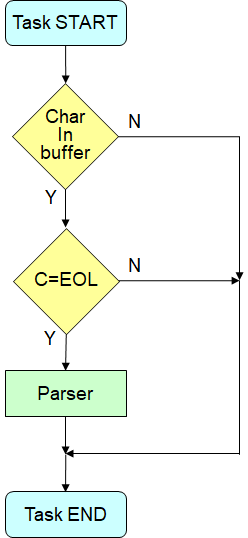}
  \caption{Flowcharts describing the EXEC and USB-communication tasks implemented in the firmware.
  } \label{fig:fwtaskscheme}
\end{figure}

The tasks to be performed are:
\begin{itemize}
\item Reading the state of the buttons (Read buttons) (every 5~ms)
\item Granting communication between the module and the PC via a USB interface (USB) (every 10~ms).
If the buffer contains a character, this is added at the end of a string variable. If this character is a line termination character, the string variable is analyzed by a subroutine (Parser) and, if it is recognized as a valid command, such command is executed.
\item Performing other operations (every 1~s).
These operations consist in testing whether the USB cable has been connected then disconnecting the battery from the system to prevent unnecessary discharge; if the P0 button is pressed during a calibration procedure, the system is switched off; the same happens after a preselected period of time following the end of the system calibration.
\item Reading the MR sensors: the period is determined by the $Conv\_ rate$ or $Cal \_ rate$ variable
\item Reading the temperature sensor (every 100~ms)
\end{itemize}
The flowcharts of two significant tasks are represented as an example in Fig.\ref{fig:fwtaskscheme}

\section{Data transfer}
The data transfer can be performed both in ASCII format (useful for debugging) and in binary format. The measurement can be executed both one by one (on demand) and continuously. In the second case, a "start-conversion" command is sent, and a continuous data flux is transferred (the DAQ rate has been previously set and can be as large as 200~Sa/s) until a "stop-conversion" command is sent.  {Under} these conditions the binary transmission is compulsory to prevent data overflow. A  {particular} transmission protocol has been designed to detect transmission errors. As seen in Sec.\ref{sec:sensor}, the data are   {two-byte signed integers}. However, the 14~bit resolution makes it impossible for some values to be generated. We use this feature to implement a one-byte transmission check. The values are transmitted after having been added to $2^{14}$ (in order to make all of them positive) and after having doubled the result (so as to make all the data values even). Under this condition, neither the most significant byte nor the least significant one can be "FF". This \textit{reserved} "FF" value is used as an end-of-line marker in the USB communication. The  data received are then truncated at the "FF" byte, and the whole data set is disregarded whenever "FF" does not occur after $K \times 3 \times 2$ significant bytes. This feature improves the system reliability, making it robust with respect to data-transfer misalignments.
\label{sec:datatransfer}

\section{Data elaboration}
The computer program that controls the device is written in LabView. It contains several units  designed to
\begin{itemize}
\item Initialize the communication
\item Select the sensor settings (dynamic range, filters, acquisition rate)
\item Check the temperature
\item Download the raw calibration data from the EEPROM
\item Analyze the calibration data and infer the conversion parameter set
\item Start the measurement
\item Convert the raw data into magnetic values
\item Analyze the magnetic data to track the magnetic source
\item Show and save the tracking output
\end{itemize}
The first operations are made at the start or on demand (particularly those described in Sec.\ref{sec:calibration}), while the last two tasks are performed online and require accurate programming to prevent data overflow with consequent  data loss or delayed system response. Particular care must be devoted to the data analysis program,  which is in charge to infer the magnet position and orientation from the magnetometric measurements. Details of this problem will be extensively provided elsewhere, while here we briefly summarize the methodologies applied to this end.

The software implemented to extract tracking data from magnetometric data is based on a standard best-fit procedure using a Levenberg-Marquardt algorithm \cite{levenberg_qam_44, marquardt_jsiam_63, meng_ijia_05_topcited, hu_ieee_10}. 
This best fit procedure inputs the sensor positions 
$ \{\vec r_k\} $ as independent variables and the magnetic fields measured as corresponding dependent variables. The position $\vec r$ of the dipole vector $\vec m$,  and the ambient field $\vec B_0$ are the fitting parameters. The latter are then estimated on the basis of a  model function considering the superposition of a homogeneous and a dipolar term:
\begin{equation}
\vec B (\Delta\vec r_k) = \frac{\mu_0}{4 \pi}\left ( 3\frac{(\vec m \cdot \Delta\vec r_k) \Delta\vec r_k}{|\Delta\vec r_k|^5}
-\frac{\vec m}{|\Delta\vec r_k|^3}
\right)+\vec B_0,
\end{equation}
where $\Delta\vec r_k=\vec r_k-\vec r$ are the relative positions of the sensors with respect to the dipole.
The fit output consists of 9 values, representing  three spatial co-ordinates of the dipole, three dipole moment components, and three background field components (we are neglecting the redundancy mentioned in the Introduction). The issues and  advantages related to possibly reducing the number of fitting parameters --in particular to avoid the redundant determination of $|\vec m|$-- will be discussed in a forthcoming paper. Here we simply note that an independent measurement of $\vec B_0$ could be performed with an additional sensor kept at a large distance from the dipole. However, this solution would require an increase of the assembly size and would reduce  the system  robustness with respect to environmental field inhomogeneities. 
The  {need to determine 9 co-ordinates renders it evident that the rule of thumb} discussed in Sec.\ref{sec:scalability} should apply for at least three 3D sensors, with obvious advantages in terms of accuracy and reliability when a larger number of sensors are close enough to detect the inhomogeneous field generated by the dipole. Assuming that the magnet moves slowly with respect to the acquisition rate, every fit output is profitably used as a guess for the next evaluation \cite{meng_ieee_10}.  As is known, a reliable guess helps  {to accelerate the convergence of non-linear functions such as} those used in the present case. 

To date, a last-step-output guess has proved to work efficiently in tests with sources moving at a speed of a few cm/s and rotating at a few rad/s. More advanced guessing, based on the analysis of a longer tracking history, could be developed for faster magnetic sources and will be assessed in future work. 

We verified that an ordinary personal computer with a single last-track guess is capable  {of running} the best-fit procedure in a time shorter than the 10~ms acquisition time, so  {as} to provide an estimation of the dipole position and orientation before  a new data set is acquired, thus   {substantially achieving} real-time functionality. 

\section{Results}
In this section, we report a
few examples of tracking results obtained with the  { system described and provide } a  
preliminary characterization of its performance. All these data are obtained with a neodymium magnet 0.5~mm in thickness and 2~mm in diameter as a dipolar source, using the prototype shown in Fig.\ref{fig:setup}. In this case (as visible in Fig.\ref{fig:setup}) the array contains 3+5 sensors on two parallel PCBs displaced by 16~mm from each other, along the $z$ direction. Hereinafter the $z$ positions  refer to the 3-sensors PCB, i.e.  the $xy$ plane is defined as that containing the 3-sensor PCB, while the 5-sensor PCB lies on the $z=16$~mm plane. The magnet can be slowly driven by means of a gear-motor to follow a circular trajectory 10~mm in radius, around the $z$ direction.

\begin{figure}[ht]
   \centering
        \includegraphics [angle=00, width= 0.8\columnwidth] {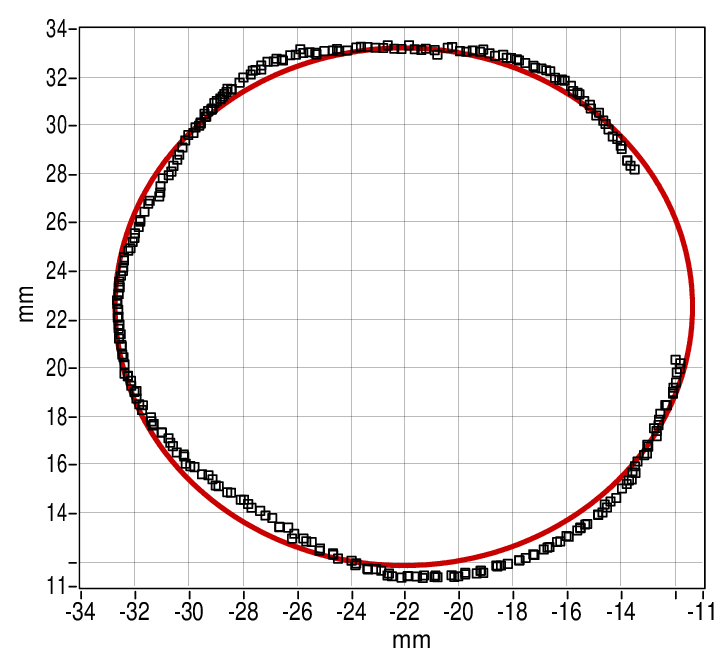}
  \caption{Position tracking 
  of a rotating magnet measured with the sensor array under static conditions. The red line shows a circular best fit of the trajectory. The acquisition rate is here set to 1/50 s.
  } 
  \label{fig:xytrackfermo}
\end{figure}

Fig.\ref{fig:xytrackfermo} shows a reconstructed trajectory as it is obtained 
with the sensor array in static condition and the dipole (which is radially oriented) that rotates on the $z= 26$~mm plane. Imperfect parallelism between  the sensor plane and the trajectory plane 
can make the trajectory projection slightly elliptic. 
However, a simplified model is sufficient to appreciate the performance of the system. With a simple circular model, we obtain a best fitting trajectory (see the red line in Fig.\ref{fig:xytrackfermo}), corresponding to a RMS error as low as 0.4~mm over a 21~mm diameter.

\begin{figure}[ht]
   \centering
        \includegraphics [angle=00,  width=0.8 \columnwidth] {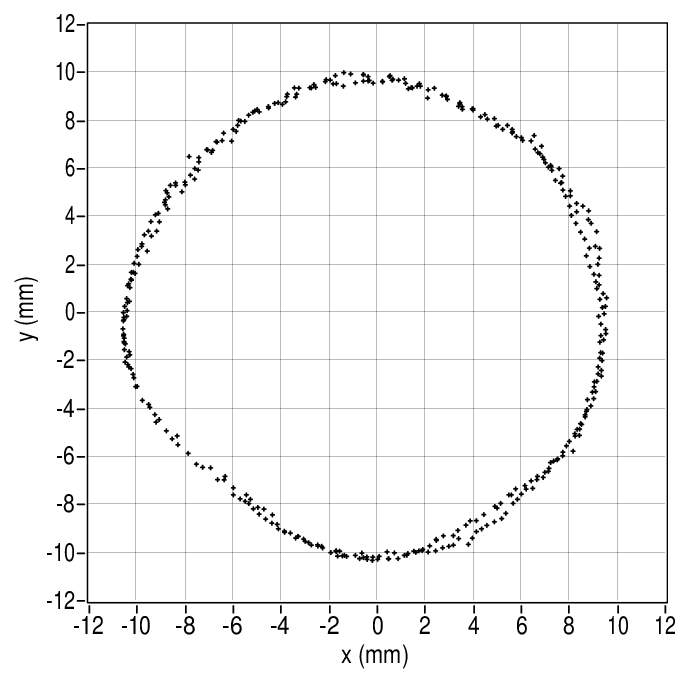}
         \includegraphics [angle=00,  width=0.92 \columnwidth] {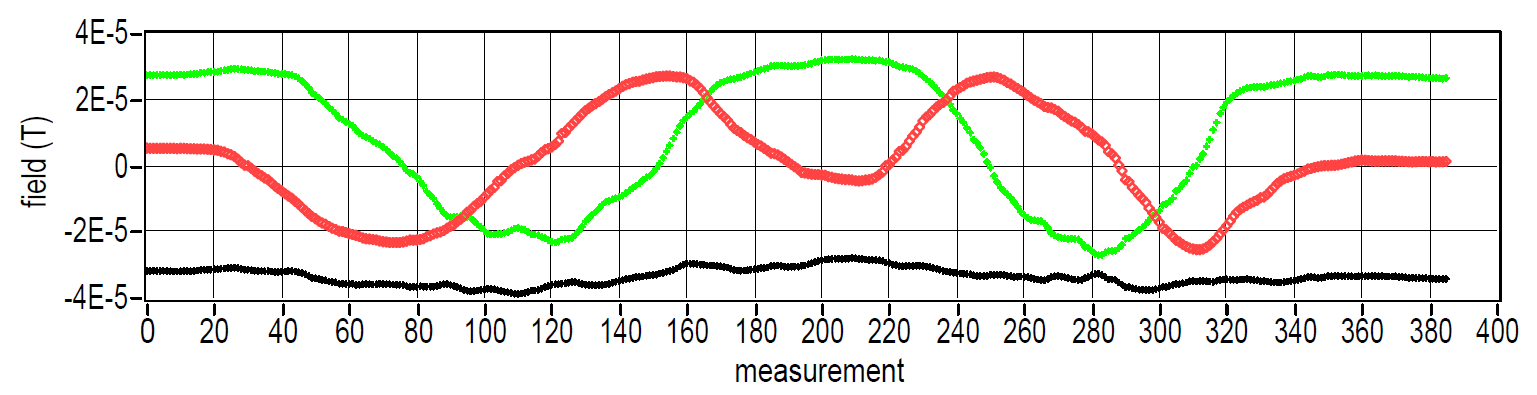}
         \includegraphics [angle=00,  width=0.9 \columnwidth] {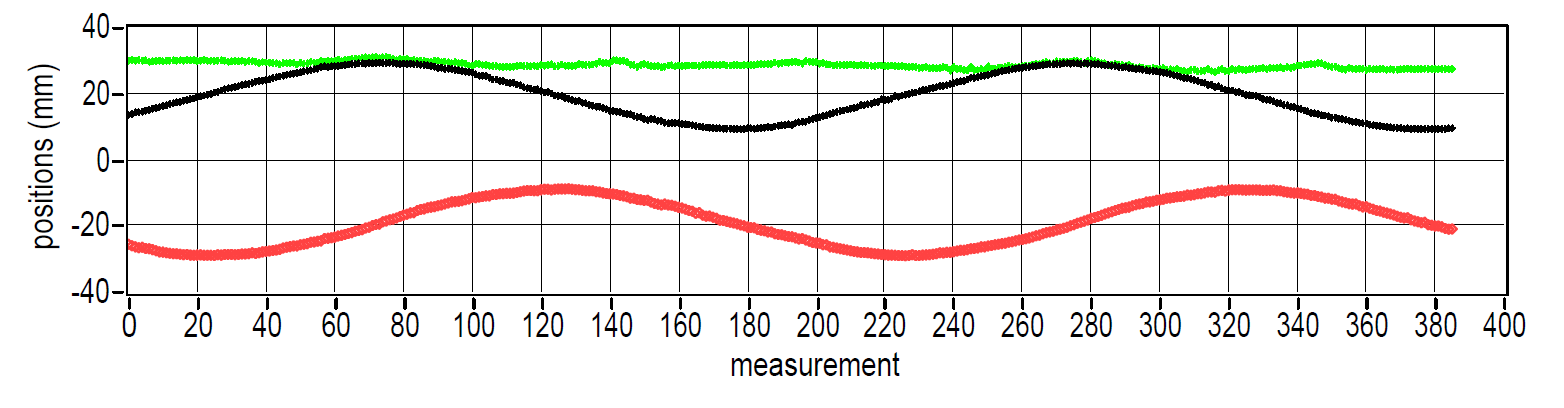}
        
  \caption{The upper plot shows the position tracking of a rotating dipole as in Fig.\ref{fig:xytrackfermo}. Here the sensor array, which is rigidly assembled with the magnet-driver, is held in the hand and   freely 
  rotated, while the magnet-driver moves the dipole with respect to the sensor array. The middle plot shows the  {three} components of the  environmental field estimated,  two of them being heavily affected by the assembly motion. The lowest plot shows the evolution of the spatial co-ordinates, with two of them following the sine/cosine law dictated by the dipole rotation and being negligibly affected by the varying orientation of the array  with respect to the environmental field.
  } \label{fig:xytrackmosso}
\end{figure}

Fig.\ref{fig:xytrackmosso}, upper 
plot, 
shows a reconstructed trajectory obtained 
with the magnet performing a similar circular trajectory on the $z=29$~mm plane. This time, the sensor array, which is rigidly assembled with the magnet driver,  is held 
in the hand and moved freely, to test the robustness of the measurement with respect to the reorientation in the environmental magnetic field. 

The  motion applied to the assembly is approximately a
rotation (about one turn backward and forward) around the $x$ axis.
In this case, the system measures the environmental field with significantly changing  components, whose  values  are represented in the middle plot of Fig.\ref{fig:xytrackmosso}, as evaluated throughout the tracking procedure. Finally, the bottom
plot shows the corresponding time evolution of the position co-ordinates. The figure proves that the reconstructed trajectory is negligibly disturbed by the array  {being in} motion. The experiment is performed in a "normal" room containing   furniture with ferromagnetic frames, computers, and other  sources of potential magnetic interference.

Preliminary quantitative estimates of the position uncertainty are based on the variance of the 3D position detected: under static conditions, 
 we measure a root-mean-square (RMS) $ \sigma_r=(\sigma^2_x+\sigma^2_y+\sigma^2_z)^{1/2}$ of 131 $\mu$m when the magnet is held still on the $z$ axis, at $z=27$~mm. Under these conditions, the relative RMS of the  environmental field measured and of the dipole modulus are  $\sigma_B/B=0.25\%$ and $\sigma_m/m=0.77\%$, respectively.

The same kind of estimation, performed when the assembly is hand-held and freely rotated, 
gives $\sigma_r=270\, \mu$m and $\sigma_m/m=1.7\%$. 
 {In short}, the position tracking is robust and sub-millimetric both when the array is kept 
static 
and when it is randomly reoriented with respect to the environmental field. The small variance of the dipole modulus estimations indicates that the system has a good  {degree of} robustness with respect to ambient field variations. In addition, the environmental field inhomogeneities typical of normal working rooms do not constitute a problem.  {In contrast, the tracking system fails when magnetized devices get too close to the sensor array.}

We notice an unexpectedly large variance in the dipole modulus estimation when the magnet moves with respect to the sensor array. For instance, 
in the case 
of the trackings in  
Figs. \ref{fig:xytrackfermo} and \ref{fig:xytrackmosso}, we achieved 
$\sigma_m/m$ values of  
about 7.6\% and 9.8\%, respectively. These large values could be ascribed  
to small uncertainties 
in the positions of  {the} sensors 
as well as  to
non-linearities in their  
responses. Despite this large variance in the dipole estimation, the quality of spatial tracking in Fig.\ref{fig:xytrackmosso} is comparable  {to that in} Fig.\ref{fig:xytrackfermo}:  {notably the greater uncertainty in the dipole vector determination is not associated with a significant degradation} of the position tracking.

\begin{figure}[ht]
   \centering
        \includegraphics [angle=270, width= \columnwidth] {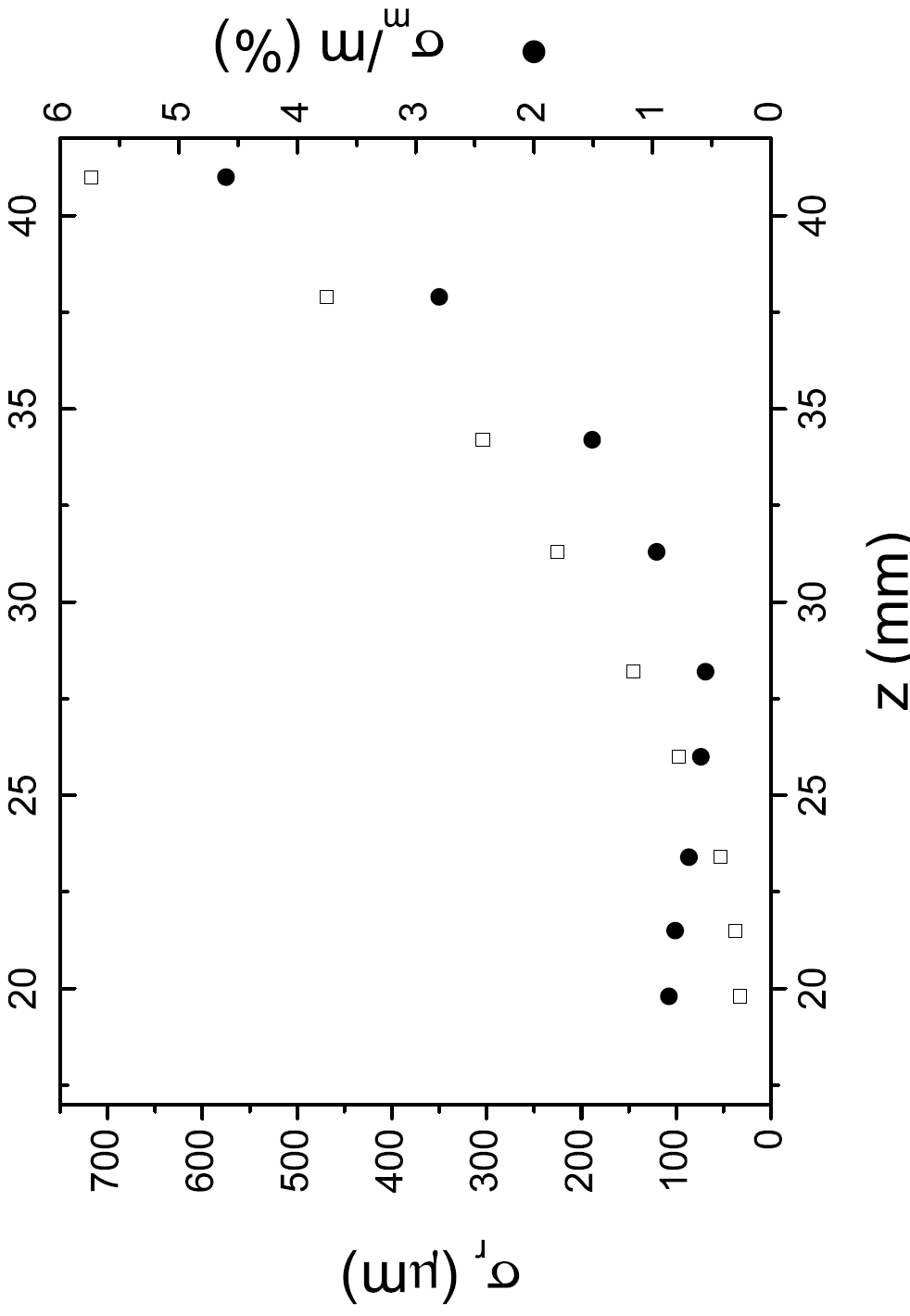}
  \caption{RMS error evaluated from repeated estimations of the dipole position  (left axis) and dipole moment modulus (right axis) as a function of the distance $z$ from the PCBs, as obtained by moving the magnet along an axis passing by center of the array.
  } \label{fig:sigme}
\end{figure}

As expected, the tracking performance with a given magnet drops dramatically when it is displaced too far from the sensor array,  {making} the dipole field comparable with the system noise, or too uniform over the array size. This behavior is  {effectively} confirmed by the data plotted in Fig.\ref{fig:sigme}. This 
figure shows the absolute position error (RMS of the reconstructed position) and the relative RMS of the dipole modulus as a function of the $z$ co-ordinate of the magnet, i.e. at increasing distances from the array.

Concerning the speed, we proved that the system can acquire and track continuously at the maximum sampling rate (currently set by the firmware at 100~Sa/s) on a standard personal computer (e.g. on an i5-7400 CPU,  3~GHz, 64~bit).

\section{Conclusion}
We have built and tested cheap and reliable hardware based on commercial magnetoresistive sensors that, after appropriate calibration procedures, provides a set of 24 magnetometric measurement data  {at a rate as high as 100~Sa/s}. The hardware contains a microprocessor enabling immediate data transfer to a personal computer, which in turns executes data elaboration to extract multidimensional (from 7D to 9D) spatial and angular co-ordinates of the magnetic source with respect to the sensor array and of the latter with respect to the ambient field. The tracker is scalable in size and may be of interest for various kinds of applications, ranging from medical diagnostics to virtual and augmented reality.

\section{Patents}
The subject of this work, in virtue of its interesting potentialities and of its demonstrated performance in terms of precision and  { speed achieved, constitutes part of the contents of a recent patent application  \cite{brevetto}.}

\vspace{6pt} 




\acknowledgments{The authors acknowledge the valuable technical support of Leonardo Stiaccini (DSFTA).}


\bibliography{bibtrack}

\end{document}